\documentclass[11pt,twoside]{article}
\usepackage{./asp2014}

\aspSuppressVolSlug
\resetcounters

\begin{document}

\title{Cold gas in High-z Galaxies: The molecular gas budget}
\author{R.~Decarli,$^{1,2}$ C.~Carilli$^{3,4}$, C.~Casey$^5$, B.~Emonts$^6$, J.A.~Hodge$^7$, K.~Kohno$^8$, D.~Narayanan$^9$, D.~Riechers$^{10}$, M.T.~Sargent$^{11}$, F.~Walter$^{2,3}$
\affil{$^1$INAF -- Osservatorio di Astrofisica e Scienza dello Spazio, via Gobetti 93/3, 40129 Bologna, Italy; \email{roberto.decarli@inaf.it}}
\affil{$^2$Max Planck Institut f\"{u}r Astronomie, K\"{o}nigstuhl 17, 69117 Heidelberg, Germany}
\affil{$^3$National Radio Astronomy Observatory, Pete V.~Domenici Array Science Center, P.O.~Box O, Socorro, NM 87801, USA}
\affil{$^4$Cavendish Laboratory, University of Cambridge, 19 J.J.~Thomson Avenue, Cambridge CB3 0HE, UK}
\affil{$^5$Department of Astronomy, The University of Texas at Austin, 2515 Speedway Blvd Stop C1400, Austin, TX 78712}
\affil{$^6$National Radio Astronomy Observatory, 520 Edgemont Road, Charlottesville, VA 22903, USA}
\affil{$^7$Leiden Observatory, Niels Bohrweg 2, 2333 CA Leiden, The Netherlands}
\affil{$^8$Institute of Astronomy, School of Science, The University of Tokyo, 2-21-1 Osawa, Mitaka, Tokyo 181-0015, Japan}
\affil{$^9$Department of Astronomy, University of Florida, 211 Bryant Space Science Center, Gainesville, FL 32611, USA}
\affil{$^{10}$Cornell University, 220 Space Sciences Building, Ithaca, NY 14853, USA}
\affil{$^{11}$Astronomy Centre, Department of Physics and Astronomy, University of Sussex, Brighton, BN1 9QH, UK}
}

\paperauthor{R.~Decarli}{roberto.decarli@inaf.it}{0000-0002-2662-8803}{INAF}{Osservatorio di Astrofisica e Scienza dello Spazio}{Bologna}{BO}{40129}{Italy}
\paperauthor{C.~Carilli}{ccarilli@nrao.edu}{}{NRAO}{Pete V.~Dominici Array Science Center}{Socorro}{NM}{87801}{USA}
\paperauthor{C.~Casey}{cmcasey@astro.as.utexas.edu}{}{University of Texas at Austin}{Department of Astronomy}{Austin}{TX}{78712}{USA}
\paperauthor{B.~Emonts}{bjornemonts@gmail.com}{}{National Radio Astronomy Observatory}{50 Edgemont Road}{Charlottesville}{VA}{22903}{USA}
\paperauthor{J.A.~Hodge}{hodge@strw.leidenuniv.nl}{}{Leiden Observatory}{}{Leiden}{CA}{2333}{The Netherlands}
\paperauthor{K.~Kohno}{kkohno@ioa.s.u-tokyo.ac.jp}{}{School of Science, University of Tokyo}{Institute of Astronomy}{Tokyo}{Mitaka}{181-0015}{Japan}
\paperauthor{D.~Narayanan}{desika.narayanan@gmail.com}{}{University of FLorida}{Department of Astronomy}{Gainesville}{FL}{32611}{USA}
\paperauthor{D.~Riechers}{riechers@astro.cornell.edu}{}{Cornell University}{}{Ithaca}{NY}{14853}{USA}
\paperauthor{M.T.~Sargent}{Mark.Sargent@sussex.ac.uk}{}{University of Sussex}{Astronomy Centre, Department of Physics and Astronomy}{Brighton}{}{BN1 9QH}{UK}
\paperauthor{F.~Walter}{walter@mpia.de}{}{MPIA}{Galaxies and Cosmology}{Heidelberg}{}{69117}{Germany}


\section{Science Goals}
The goal of this science case is to accurately pin down the molecular gas content of high redshift galaxies. By targeting the CO ground transition, we circumvent uncertainties related to CO excitation. The ngVLA can observe the CO(1-0) line at virtually any $z>1.5$, thus exposing the evolution of gaseous reservoirs from the earliest epochs down to the peak of the cosmic history of star formation. The order-of-magnitude improvement in the number of CO detections with respect to state-of-the-art observational campaigns will provide a unique insight on the evolution of galaxies through cosmic time.


\section{Scientific rationale}

Our understanding of the evolution of galaxies, and in particular of the cosmic history of star formation, is limited by our ignorance of the amount of cold gas that galaxies could use to fuel star formation in various cosmic epochs. The bulk of the Universe's cold gas reservoir is comprised of hydrogen gas in the form of atomic hydrogen, H{\sc i}, and molecular hydrogen, H$_2$, the latter of which is responsible for star formation in cold, condensed molecular clouds. A key goal of the ngVLA will be to assess the molecular gas content in high-redshift galaxies.

\begin{figure}
\centering
\includegraphics[width=0.99\textwidth]{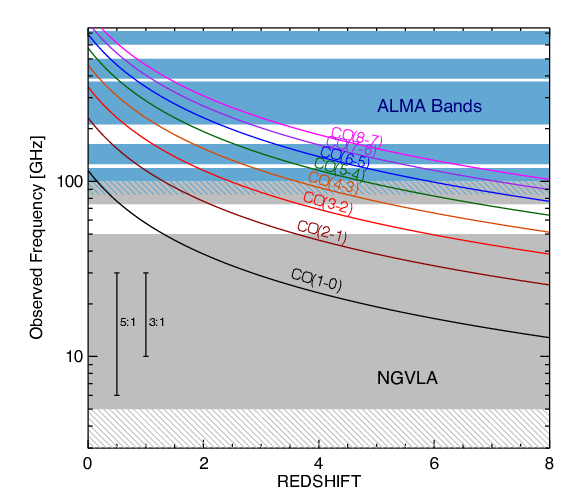}
\caption{The observed frequencies of various transitions of CO from $z$=0 to $z$=8.  While the ALMA bands (blue shading) are sensitive to many transitions of CO at low redshift, the low-J CO lines redshift out of the bands beyond $z\sim0.5-2$. These low-J CO lines are critical for characterization of the total molecular gas reservoir in high-$z$ galaxies and are only currently accessible for high-redshift galaxies using the VLA. The gray region indicates the ngVLA coverage of 5--100 GHz, with a gap around the 59--75 GHz water band. The gray hashed region represents the reach capabilities for ngVLA at low frequencies. With the advent of very wide-bandwidth setups for the ngVLA, we will be able to obtain multiple CO transitions for galaxies beyond $z\sim3.5$ in one correlator setup, which will be revolutionary for studying dust and gas in very high-redshift galaxies. Figure adapted from \citet{casey15}.}
\label{fig_wg3_mh2_freqrange}
\end{figure}

Although abundant, H$_2$ is not directly observable under normal circumstances in the star-forming medium. The bright transitions of carbon monoxide (CO), the second most abundant molecule in the universe, are thus the workhorse for investigating the molecular content of galaxies through cosmic time. Indeed, the vast majority of the few hundred molecular line detections reported so far at $z>1$ have been of CO (see more in reviews of \citealt{solomon05}, \citealt{carilli13}). Due to frequency shifting at sufficiently high-redshift, millimeter-operating facilities like IRAM's PdBI/NOEMA, CARMA, and ALMA can only observe the higher-J transitions of CO. The intrinsic brightness of these lines depends not only on the amount of molecular gas, but primarily on the excitation of the CO molecules, which is affected by the temperature and density of the gas. The so-called CO Spectral Line Energy Distribution (SLED) might vary substantially in different regions of a galaxy by up to factors of five, and even between the relatively low-J transitions from CO(3-2) to CO(1-0) --- this implies that high-J transitions cannot be used to gauge the molecular gas mass. On the other hand, the lower-J transitions, including the ground state CO(1-0) transition, are only accessible to longer wavelength radio observatories like the VLA (see Figure \ref{fig_wg3_mh2_freqrange}).

From the CO luminosity one can infer molecular gas masses $M_{\rm H2}$ via the CO--to--H$_2$ conversion factor, $X_{\rm CO}$ or $\alpha_{\rm CO}$. This conversion varies by up to a factor of five depending on the conditions of the gas in the ISM, including metallicity. Recent work has focused on using dust continuum measurements to scale to gas masses \citep{magdis11,magdis12,santini14,scoville14,scoville16}. However,  this technique typically assumes that the dust-to-gas ratio is fixed across a wide range of galaxies and redshifts, which has been shown to be a poor assumption  in  some  cases  \citep{remyruyer14,capak15}. In particular, the dust-to-gas ratio is known to depend strongly  on metallicity  \citep[e.g.,][]{issa90,lisenfeld98,draine07,bolatto13,berta16}. Furthermore, the important kinematic signatures that come along with molecular line measurements are missed in dust continuum measurements.

\begin{figure}
\centering
\includegraphics[width=0.99\textwidth]{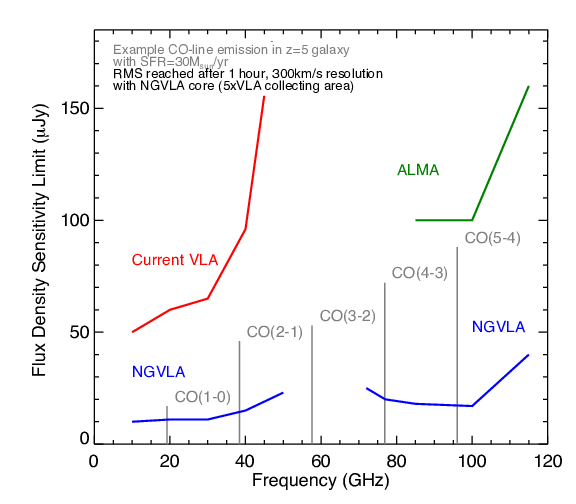}
\caption{The projected sensitivity limits of the proposed ngVLA core, which will have five times the effective collecting area of the current VLA. This plot also shows the sensitivity limits of ALMA alongside the current VLA and projected ngVLA as a function of frequency (the gap from 50--75 GHz is due to atmosphere). Overplotted are the anticipated CO line strengths, from CO(1-0) to CO(5-4) of a typical z=5 galaxy (with SFR=30 M$_\odot$\,yr$^{-1}$). This type of galaxy is currently well below the detection limit of existing facilities, including ALMA, yet is crucial for understanding the star forming budget of the Universe's first galaxies. Figure adapted from \citet{casey15}.}
\label{fig_wg3_mh2_sensitivity}
\end{figure}

The ideal probe of the cold molecular gas reservoir is therefore the ground state transition of CO, CO(1-0),  where the scaling to molecular hydrogen mass has only the uncertainty in $X_{\rm CO}$ with which to contend. Constraining $X_{\rm CO}$ in high-$z$ galaxies is possible via empirical relations linking the conversion factor with the dynamical mass and stellar mass, which further reduce the uncertainty on gas mass. The JVLA, GBT, and  ATCA have detected CO(1-0) in a number of high-redshift starburst galaxies, albeit a few at a time with long integrations \citep{ivison11,hodge12,papadopoulos12,greve14,emonts14}. These initial detections of CO(1-0) at high-redshift have not only constrained the total molecular gas mass of hydrogen in these high-$z$ systems, but also highlighted the importance of understanding the spatial distribution of gas in distant galaxies. While  the  submillimeter-luminous galaxy population, i.e., dusty star forming galaxies (DSFGs) have been found to be relatively compact in high-J transitions \citep{tacconi08,bothwell10}, consistent with the idea that they are scaled-up analogs to local ULIRGs \citep{sanders96}, their CO(1-0) maps are dramatically different, showing gas extending as far as 16 kpc from the galaxy centers (\citealt{ivison11,spilker15}; see also \citealt{decarli16b}). Studies of high-redshift proto-cluster radio galaxies even revealed bright CO(1-0) reservoirs out to distances of $\sim$60 kpc from the central galaxy \citep{emonts14}. While these larger sizes are more suggestive of disk-like rotation dynamics and widespread molecular gas reservoirs in the halo environment  of  massive  galaxies,  the number  of  galaxies surveyed at the current VLA sensitivity makes it difficult to infer large-scale population dynamics or draw comparisons between low-$z$ and high-$z$ galaxy populations --- a situation that will not significantly change even after ALMA starts to offer band 1\&2 observations. The ngVLA is desperately needed to dramatically increase the number of galaxies surveyed in CO(1-0) at high-redshift --- by factors in the thousands, in line with what is currently observed from high-redshift galaxies via direct starlight. Figure \ref{fig_wg3_mh2_sensitivity} illustrates the anticipated improvement in ngVLA sensitivity over the current VLA and ALMA depths across 10--120 GHz.

To demonstrate the dramatic impact of ngVLA on detection rates of CO(1-0) in high-$z$ samples, we generated a mock observation with a 3:1 bandwidth ratio tuned to 11--33 GHz. This frequency range will pick up the CO(1-0) transition from $2.5<z<9.5$ {\em in a single frequency setting}. The RMS reached after a one hour integration with five times the collecting area of the current VLA is $\sim10$ $\mu$Jy\,beam$^{-1}$ RMS in 2 MHz bins, sufficient to resolve individual CO lines into 3--5 spectral bins across the entire bandwidth. 

\begin{figure}
\centering
\includegraphics[width=0.99\textwidth]{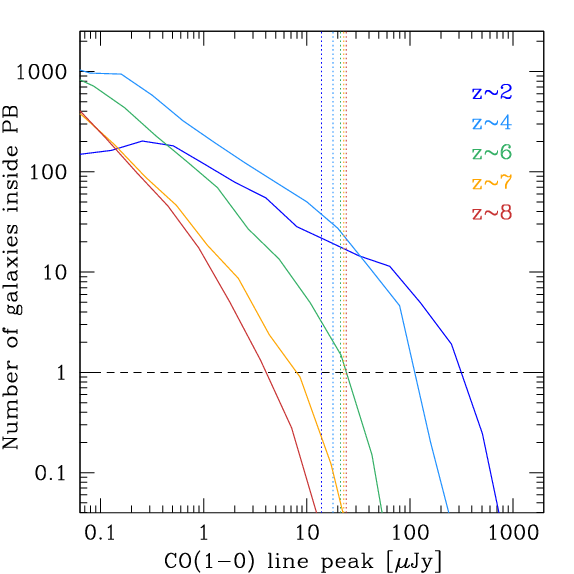}
\caption{Predicted CO(1-0) flux distributions in various redshift bins, according to the semi-analytical models by \citet{popping12,popping14,popping16}. The vertical lines mark the sensitivity that we will reach with ngVLA in the example of molecular deep field outlined in the text. Tens of galaxies at $z=2-4$ will be detected within a single pointing. Recent molecular deep field investigations (ASPECS: \citealt{decarli16a}, 2018 in prep.; COLDz: Riechers et al.~2018 in prep.) suggest that semi-analytical models tend to underpredict the bright end of the CO luminosity function, i.e., the expected detection counts per ngVLA pointing might exceed 100. For a comparison, at the present sensitivity only a handful of detections per pointings are expected.}
\label{fig_wg3_mh2_predictions}
\end{figure}

To estimate the number of galaxies accessible to CO(1-0) detection within one ngVLA pointing in a 11--33 GHz frequency scan, we computed the flux distribution of CO emitters expected based on state-of-the-art semi-analytical models of galaxy formation and evolution \citep{popping12,popping14,popping16}. These models couple a semi-analytic model of galaxy formation with a radiative transfer code to make predictions for the luminosity function of CO. 
They treat the interstellar medium as a density distribution for each galaxy, and assume a log-normal density distribution for the gas within clouds. They then consider only the UV contribution to the heating (no X-rays), and model the CO chemistry based on a fit to results from the photodissociation region code of \citet{wolfire10}. The value of $\alpha_{\rm CO}$ is left free to vary from galaxy to galaxy based on the various galaxy properties (mass, compactness, SFR, etc). We transform the luminosity function of CO(1-0) computed at various redshifts into flux distributions, focusing on the redshift interval $z>2.5$. We consider the volume corresponding to the ngVLA primary beam (assuming 25m antennas) integrated along the line of sight in various contiguous redshift slices. The total volume of the universe sampled with such a scan is $\sim$88,500 comoving Mpc$^3$. By scaling the CO luminosity / flux distributions to this volume, we obtain the expected number of CO-emitting galaxies at various redshifts in our spectral scan. If we restrict our search to the limit of $\sim10$ $\mu$Jy\,beam$^{-1}$ RMS in 2 MHz bins, we consider a typical line width of 200 km\,s$^{-1}$, and we require secure $>5$-$\sigma$ detections, we expect 33 CO(1-0) detections at $z\sim2$; 43 at $z\sim4$; and even 2 at $z\sim6$ (see Figure \ref{fig_wg3_mh2_predictions}). To first order, these estimates are in general agreement with the relatively coarse observational constraints on the CO luminosity functions at high redshift available so far \citep{walter14,decarli16a} and with predictions based on recent estimates of the integrated infrared luminosity function \citep{casey14,vallini16} which is roughly complete out to $z\sim 2$, translated into a best-guess CO(1-0) luminosity function using a conservative estimate of a ULIRG-type $L'_{\rm CO(1-0)}$-to-$L_{\rm IR}$ scaling \citep{bothwell13} and assuming a line width of 200 km\,s$^{-1}$. Further support to these estimates is derived from the empirical predictions for the number of line emitters in a CO deep field by \cite{dacunha13a}, based on a physically motivated spectral energy distribution model along with empirical relations between the observed CO line and infrared luminosities of star-forming galaxies. These estimates capitalized on the deepest available optical/near-infrared data for the Hubble Ultra Deep Field (UDF). Given all of the assumptions that went into the models, and particularly since the estimates were derived in very different ways, this relative agreement is encouraging. Intriguingly, preliminary results from larger campaigns blindly searching for molecular gas emission in deep fields suggest that the predictions based on semi-analytical models might underestimate the number of CO--bright sources at the high--luminosity end of the luminosity functions (see results from ASPECS LP in Decarli et al.~2018, in prep; and from COLDz in Riechers et al.~2018, in prep). If these preliminary results are confirmed, the expected number of CO detections per ngVLA pointings might exceed 100.

The mock observation described in this paper will not only be sensitive to CO(1-0), but also CO(2-1) emission at early epochs. However, the sensitivity and field of view for CO(2-1) will be significantly different for the same redshift regime given the differences in frequency, and will more closely mirror the coverage for the low-redshift CO(1-0) emitters. The survey sensitivity with respect to CO(2-1) depends on source excitation as the ratio of $S_{\rm CO(2-1)}$=$S_{\rm CO(1-0)}$ can vary from 2--4. Perhaps counter-intuitively, the depth of CO(2-1) observations at high-$z$ will be greater than for CO(1-0), although the number of expected sources per single pointing will not differ substantially from the expected CO(1-0) emitters given the limited field of view. Line degeneracy may then be of concern, since it will be difficult to disentangle low-redshift CO(1-0) emitters and high-redshift CO(2-1) emitters.  Other multiwavelength information (and perhaps parallel ALMA observations) would be necessary to break this degeneracy. 

The estimates discussed so far refer to an individual pointing with 25m antennas. Clearly, smaller antenna dishes would correspond to an increased surveyed area, and therefore larger number of detections at a given survey depth. In case of a mosaic, we note that weaving a series of single pointings together will have unique advantages with such a wide bandwidth. For a single pointing example above, the low frequency space probes the largest volumes, but at shallower luminosity limits at high-redshifts. Tiling a mosaic together by maximizing overlap for the low-frequency observations will dramatically reduce the RMS at higher-redshifts, while mapping a large area. Figure 4 shows the effect of  mosaicing on the CO(1-0) luminosity limit out to high-redshifts, sensitive to objects in the gray parameter space. 

\begin{figure}
\centering
\includegraphics[width=0.99\textwidth]{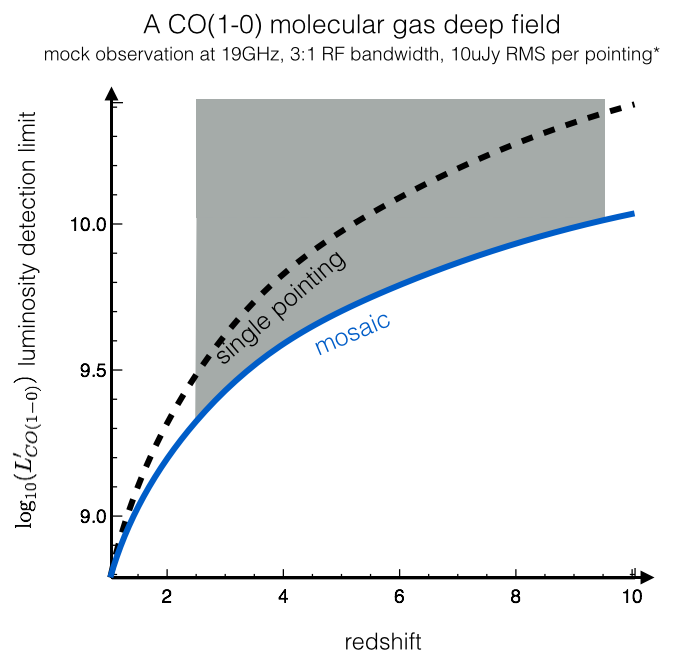}
\caption{Luminosity limit per pointing reached in a molecular deep field initiative, with a single pointing and with a mosaic. By tiling different pointings with appropriate overlap, we significantly improve on the final depth of the data in any given region. Figure adapted from \citet{casey15}.}
\label{fig_wg3_mh2_co_lf}
\end{figure}

While direct molecular gas detections are certainly the main focus of future ngVLA extragalactic deep fields, it is worth noting that these surveys will be done in well-surveyed legacy fields, where there is already substantial ancillary data including spectroscopic redshifts.  With large numbers of spectroscopic redshifts, stacking to measure the median molecular gas reservoir in LBGs would become possible as a function of redshift, environment, or other physical factors.

An important caveat is that, at sufficiently high redshift, the temperature of the CMB is non-negligible and can contribute to dust continuum and gas heating within galaxies. While this extra heating of the gas (and dust) will boost the intrinsic fluxes, the CMB also becomes a stronger background against which these fluxes must be measured. The net effect of these two competing effects on CO emission is less straightforward than for the dust continuum emission \citep{dacunha13b}, as the regions of cold molecular gas may not be in local thermal equilibrium, or the kinetic temperature of the gas may not be thermally coupled to the dust in an obvious way, particularly in early-Universe galaxies.  However, the overall effect of CMB heating will be to cut down the number of direct detections in the $z\gtrsim6$ Universe at low gas temperatures ($T\lesssim 20$ K). 

On a technical note, using phased array feeds (PAFs) with multiple pixels instead of single pixel feeds would allow the ngVLA to survey areas of sky corresponding to volumes of $\sim$0.1 Gpc$^3$, covering the full range of cosmic environments.  For example,  if we were to use a $3 \times 3$ pixel PAF we could survey the entire 2 deg$^2$ COSMOS field to this depth in a few hundred hours, detecting $\sim10^5$ CO emitting galaxies. This would allow us, for example, to compute 2-point correlation functions within the CO emitting population, and cross-correlations between CO emitting galaxies and other populations such as galaxies with large stellar masses, galaxy clusters and AGN, all of which can be compared directly with  the output of simulations. This is critical for assessing the collapse of large scale structure and environmental impact on star formation in galaxies. 

\subsection{Measurements Required}

The science goals presented here require the detection of faint lines distributed over large volumes. The key features will be the sensitivity to faint lines, and the wide simultaneous coverage in terms of both field of view and redshift (i.e., frequency) space.

\subsection{Uniqueness to ngVLA Capabilities (e.g., frequency coverage, resolution, etc.)}
The key features needed in order to successfully address the proposed science goals are:\\
1) sensitivity to faint line observations in the 12--50 GHz range, in order to sample CO(1-0) at the highest redshifts;\\
2) imaging capabilities to 1--3$''$ scales, in order to unambiguously identify counterparts without out-resolving the faint line emission;\\
3) large bandwidth: $\Delta \nu/\nu>1$, in order to maximize the simultaneous CO redshift coverage and therefore the survey speed;\\
4) large field of view: $>1$ arcmin$^2$ per pointing at any frequency, again to insure rapid coverage of large volumes.
The combination of requirements 1 and 2 forces us to restrict to radio interferometers. The VLA in principle has the power to achieve these goals, but lacks the required sensitivity (see Figure \ref{fig_wg3_mh2_sensitivity}). ALMA performs better in terms of sensitivity at $\sim$3 millimeter wavelengths, but it lacks the frequency coverage below 80 GHz, which is crucial to observe CO(1-0) at high redshift.

\subsection{Longevity/Durability: with respect to existing and planned ($>$2025) facilities}

The molecular scan efforts will inevitably focus on well studied fields where exquisitely deep observations at various wavelengths are already in place, and more will come in the coming years. E.g., all the key extragalactic fields (HDF-N, HUDF, GOODS, COSMOS, EGS, etc) have been heavily hammered with HST, Spitzer, Herschel, ALMA, Chandra, XMM, VLT, Keck, and will be natural targets for extensive investigations with upcoming facilities and observatories, such as JWST, WFIRST, TMT/E-ELT. Optical/NIR images will be used to characterize the stellar populations in the galaxies detected in CO(1-0) in the ngVLA molecular scans. Optical/NIR spectroscopy will provide precise redshifts that can be used to further push the sensitivity of the ngVLA observations via stacks. For the brightest sources, optical/NIR spectra will also provide information on, e.g., the gas metallicity and other properties of the ISM thanks to rest-frame optical diagnostics. X-ray observations will allow to separate AGN from quiescent galaxies, thus allowing us to put the ngVLA measurements into the broader context of galaxy and black hole joint evolution.  The multi-wavelength photometric coverage will provide precise stellar masses and star formation rates, which will be crucial to complement the molecular gas information provided by ngVLA. JWST data will enormously improve on the existing data, both in terms of wavelength coverage, sensitivity, and angular resolution (especially in the MIR), and in terms of spectroscopic capabilities at NIR and MIR wavelengths. Similarly, the advent of 30 m class telescopes will open up the possibility of spectroscopically characterizing sources that are barely detected in broad-band photometry with the state-of-the-art instrumentation --- again fostering an obvious synergy with the ngVLA effort.

\acknowledgements ...  



\end{document}